\documentclass[manuscript]{acmart}

\usepackage{stfloats}
\usepackage{booktabs}

\copyrightyear{2025}
\setcopyright{cc}

\def\authnote{0} 
\def\editsview{0} 
\def\revview{0} 
\newcounter{mynote}[section]
\newcommand{\notecolor}{blue}
\newcommand{\thenote}{\thesection.\arabic{mynote}}
\newcommand{\tnote}[1]{\ifnum\authnote=1\refstepcounter{mynote}{\bf \textcolor{\notecolor}{$\ll$TomR~\thenote: {\sf #1}$\gg$}}\fi}

\renewcommand{\paragraph}[1]{\vspace*{6pt}\noindent\textbf{#1}\;}

\newcommand{\fixme}[1]{\ifnum\authnote=1{\textcolor{red}{[FIXME: #1]}}\fi}
\newcommand{\better}[1]{\ifnum\authnote=1{\textcolor{violet}{[BetterWord: #1]}}\fi}
\newcommand{\todo}[1]{\ifnum\authnote=1{\textcolor{red}{[TODO: #1]}}\fi}

\newcommand{\edit}[1]{\ifnum\editsview=1{\textcolor{red}{#1}}\else{#1}\fi}

\newcommand{\rev}[1]{\ifnum\revview=1{\textcolor{red}{#1}}\else{#1}\fi}

\newcommand{\dnote}[1]{\ifnum\authnote=1\refstepcounter{mynote}{\bf \textcolor{red}{$\ll$DA~\thenote: {\sf #1}$\gg$}}\fi}

\newcommand{\ynote}[1]{\ifnum\authnote=1\refstepcounter{mynote}{\bf \textcolor{red}{$\ll$YY~\thenote: {\sf #1}$\gg$}}\fi}

\newcommand{\tabitem}{~~\llap{\textbullet}~~}

\begin{document}

\title[Designing Technologies for Value-based Mental Healthcare]{Designing Technologies for Value-based Mental Healthcare: Centering Clinicians' Perspectives on Outcomes Data Specification, Collection, and Use}

\author{Daniel A. Adler}
\email{daa243@cornell.edu}
\orcid{0000-0003-3328-0312}
\affiliation{%
  \institution{Cornell Tech}
  \country{USA}
}

\author{Yuewen Yang}
\email{yy2228@cornell.edu}
\orcid{0000-0001-5490-314X}
\affiliation{%
  \institution{Cornell Tech}
  \country{USA}
}

\author{Thalia Viranda}
\email{tv74@cornell.edu}
\orcid{0009-0008-6933-3926}
\affiliation{%
  \institution{Cornell Tech}
  \country{USA}
}

\author{Anna R. Van Meter}
\email{anna.vanmeter@nyulangone.org}
\orcid{0000-0003-0012-206X}
\affiliation{%
  \institution{NYU Grossman School of Medicine}
  \country{USA}
}

\author{Emma Elizabeth McGinty}
\email{emm4010@med.cornell.edu}
\orcid{0009-0003-5095-9308}
\affiliation{%
  \institution{Weill Cornell Medicine}
  \country{USA}
  }

\author{Tanzeem Choudhury}
\email{tanzeem.choudhury@cornell.edu}
\orcid{0000-0002-5952-4955}
\affiliation{%
  \institution{Cornell Tech}
  \country{USA}
}

\renewcommand{\shortauthors}{Adler et al.}

\begin{abstract}
Health information technologies are transforming how mental healthcare is paid for through value-based care programs, which tie payment to data quantifying care outcomes.
But, it is unclear what outcomes data these technologies should store, how to engage users in data collection, and how outcomes data can improve care.
Given these challenges, we conducted interviews with 30 \rev{U.S.-based} mental health clinicians to explore the design space of health information technologies that support outcomes data specification, collection, and use in value-based mental healthcare.
Our findings center clinicians' perspectives on \rev{aligning outcomes data for payment programs and care}; opportunities for health technologies and personal devices to improve data collection; and considerations for using outcomes data to hold stakeholders including clinicians, health insurers, and social services financially accountable in value-based mental healthcare.
We conclude with implications for future research designing and developing technologies supporting value-based care across stakeholders involved with mental health service delivery.
\end{abstract}

\ccsdesc[300]{Human-centered computing~User studies}
\ccsdesc[500]{Applied computing~Health care information systems}
\ccsdesc[100]{Human-centered computing~Ubiquitous and mobile devices}

\keywords{health information technology; mental health; user-centered design; health services; implementation science; value-based care; qualitative research; passive sensing; digital phenotyping; digital biomarkers; digital mental health}

\maketitle

\section{Introduction}
\label{sec:intro}

Health information technologies (HITs) are transforming how health services collect, share, and use data.
Electronic health records (EHRs) collect clinical data on provided treatments and patients' health, which can be aggregated and shared with regulators \cite{birkhead_uses_2015}.
Personal devices, such as smartphones and wearables, create opportunities to bring data on behavior, physiology, and well-being from everyday life into clinical care \cite{mohr_personal_2017, torous_new_2017}.
These data streams are being repurposed to transform how we pay for health services through \textit{value-based care} (VBC) programs, where healthcare providers (eg, hospitals, clinicians) are paid based upon the ``value'' of care they deliver to patients \cite{world_economic_forum_moment_2023, lewis_value-based_2023}.
VBC programs are implemented by paying providers for effectively managing patients' health, instantiated by collecting data to quantify the \textit{quality of care} providers deliver \cite{world_economic_forum_moment_2023, donabedian_quality_1988}. 
VBC is well-motivated: healthcare spending is on the rise -- \rev{for many reasons, including increased prescription drug and medical device costs, increased service utilization, and a global aging population \cite{martin_national_2023, stucki_what_2023} -- but increased spending is not always associated with improved health outcomes \cite{gunja_us_2023}}.
VBC programs incentivize healthcare providers to deliver treatments that simultaneously improve health while reducing service utilization and cost \cite{mcclellan_improving_2017}. 
But, these new financial arrangements raise a variety of sociotechnical questions.
For example, what data quantifies the quality of care, and how should this data be collected and managed?
How should quality data hold health systems accountable to improve patient outcomes?
How do we design HITs that support this process?

In this work, we explored these questions in a specific area of healthcare with longstanding quality challenges: mental healthcare.
The prevalence of mental health disorders and need for treatment continues to rise \cite{wolf_scoping_2024, santomauro_global_2021, xue_mental_2024, mcbain_mental_2023}.
Despite the need for high quality mental health services, there is a large gap between evidence-based practices and delivered care \cite{institute_of_medicine_improving_2006}, and mental healthcare has been much slower to improve services compared to other healthcare specialties \cite{pincus_quality_2016}.
A recent review paper found that only 27\% of published clinical reports describe adequate adherence to mental health clinical practice guidelines \cite{bauer_review_2002}.
Data from the United States suggests that only half of publicly insured patients receive appropriate follow-up care after a mental health-related emergency department visit \cite{the_national_committee_for_quality_assurance_follow-up_2022}, \rev{and there are an inadequate number of inpatient psychiatric beds and/or care providers available to treat patients \cite{american_psychiatric_association_psychiatric_2022, kaiser_family_foundation_mental_2024}.
Simultaneously, some psychiatric hospitals keep patients in care longer then medically necessary \cite{silver-greenberg_how_2024}.}
Researchers and policymakers have proposed that VBC programs could incentivize health systems to reduce these quality gaps by creating financial incentives to improve patient outcomes \cite{hobbs_knutson_driving_2021}.
Given these challenges, in 2021, the National Committee for Quality Assurance, or NCQA -- a leading organization in the United States responsible for assessing care quality -- proposed a quality measurement framework for value-based mental healthcare \cite{niles_behavioral_2021}.
With this proposal, the NCQA placed a shared responsibility on policymakers, health insurance companies, and healthcare providers to coordinate the definition, measurement, and management of quality.
Through systems of \textit{joint accountability}, the NCQA recognized that these different stakeholders must work together to improve mental health outcomes.
But, the NCQA's proposal stopped short of defining what mental health outcomes these programs should use, how outcomes data should be collected, and how accountability should be shared.

We therefore studied the design of HITs that support outcomes data storage, collection, and use in VBC with an important stakeholder in this process: mental health providers, specifically practicing clinicians.
We focused this study on mental health clinicians since they will play an essential role in realizing VBC. 
Mental health clinicians use their expertise to decide what treatments patients receive, they collect data on patients' progress in treatment, and collected data is transformed into the quality metrics that will determine how clinicians are reimbursed for their services.
Thus, VBC holds clinicians financially accountable to make treatment decisions based upon specific quality metrics, and assumes that improving metrics will improve patients' health.
Mental health clinicians have found these new forms of accountability challenging.
For example, less than 20\% of mental health clinicians practice \textit{measurement-based care} -- the process of routinely collecting data to measure treatment outcomes and inform decision-making \cite{fortney_tipping_2017, kilbourne_measuring_2018} -- citing concerns that data collection is burdensome, data collected within MBC do not effectively measure care outcomes across patients \cite{tauscher_what_2021, barkham_routine_2023}, and data could be used punitively to influence clinicians' pay \cite{lewis_implementing_2019, desimone_impact_2023}. 
These tensions create opportunities to study with mental health clinicians what data better measure care outcomes, how this data should be collected, and the extent to which data could be repurposed to create more accountable care.

In this work, we contribute findings from a set of interviews with 30 \rev{U.S.-based} mental health clinicians to explore the design space for HITs that support (1) outcomes data specification, (2) collection, and (3) use as a part of value-based mental healthcare.
These three areas were inspired by Li et al's \textit{stage-based model of personal informatics} \cite{li_stage-based_2010}, specifically the stages of \textit{preparation, collection, and action}, applied in this work to study the HITs and people (mental health clinicians) that will prepare, collect, and use outcomes data in value-based mental healthcare.
Our findings center mental health clinicians' perspectives in the design of HITs supporting value-based care.
Specifically, participating clinicians advocated that HITs store functional and engagement outcomes for value-based mental healthcare, which were perceived as the proximal outcomes to provided services (Section \ref{sec:findings:preparation}); called for investments in HITs that reduce the burden of collecting outcomes data (Section \ref{sec:findings:collection}); and believed that outcomes data would need to hold providers, health insurers, and social services jointly accountable to improve care (Section \ref{sec:findings:action}).
We conclude with implications for research developing (Section \ref{sec:discussion:tech}) and designing (Section \ref{sec:discussion:design}) HITs to better align stakeholders' -- including payers, clinicians, and social services -- data needs for both VBC and patient care.

\section{Related Work}
\label{sec:rw}

Our work lies at the intersection of three lines of inquiry: research on technologies supporting health services (Section \ref{sec:rw:tech-services}), mental health data collection and storage (Section \ref{sec:rw:data}), and value-based mental healthcare (Section \ref{sec:rw:vbc}).

\subsection{Designing Technologies for Health Services}
\label{sec:rw:tech-services}

In this work, we studied technologies that support value-based care and the delivery of \textit{health services}, which encompass the people, organizations, and technology involved in healthcare delivery \cite{issues_working_1994, sanford_schwartz_chapter_2017}.
These people and organizations include \textit{healthcare providers}, the clinicians or hospital systems that provide treatments or preventive care (the ``services''); as well as \textit{healthcare payers}, the government agencies or private health insurance companies that pay for health services.
We review specific technologies supporting mental health services in Section \ref{sec:rw:data}.
To design technologies for health services, we need to confront more than the hardware or software capabilities of a specific technology, or the effectiveness of interventions that use technologies to improve health outcomes.
We also need to confront sociotechnical factors that affect the implementation and effectiveness of these technologies in real-world care. 
Norman and Stappers categorize sociotechnical factors that affect technology implementation as political, economic, cultural, organizational, and structural \cite{norman_designx_2015}.
Blandford states that, for health services specifically, HCI scholars should \textit{``consider stages (of identifying technical possibilities or early adopters and planning for adoption and diffusion) that are rarely discussed in HCI, but that are necessary to deliver real impact from HCI innovations in healthcare''} \cite{blandford_hci_2019}.
Thus, we were motivated to improve the design of technologies supporting health services by understanding factors that affect their implementation and adoption in care.

Recently, HCI scholars have considered adopting ideas from health services research to improve both the design and effectiveness of health technologies.
Scholars have considered how HCI research can integrate aspects of \textit{implementation science} -- the health services field examining the real-world adoption of evidence-based interventions \cite{lyon_bridging_2023}. 
Interviews with HCI and implementation science researchers uncovered that HCI tends to de-prioritize factors that influence long-term adoption of technologies in their initial design, including the financial incentives that affect adoption, and an understanding of how technologies support providers after implementation \cite{dopp_aligning_2020}.
Moreover, HCI scholars have stated that if technologies are to impact real-world care, HCI researchers should focus on how technology is consumed in care, including developing an understanding of the technical and market incentives to use new tools \cite{colusso_translational_2019}.
Inspired by this work, we considered these aspects of adoption in the initial design of technologies that support value-based mental healthcare.
Specifically, we considered how technologies can support healthcare providers -- practicing clinicians -- including how these technologies can be integrated into clinicians' workflows to support care, and the financial incentives that influence HIT adoption as a part of value-based care.

\subsection{Health Information Technologies for Collecting and Storing Mental Health Data}
\label{sec:rw:data}

HCI, health informatics, and mental health researchers have collaborated to build health information technologies (HITs) for collecting and storing mental health data.
In this work, we focus on three categories of mental health data: clinical data, active data, and passive data.
\textit{Clinical data} can be retrieved from \textit{electronic health records} (EHRs), which record information collected during clinical visits including patient demographics, diagnoses, health and family history, treatments provided, and unstructured clinical notes \cite{birkhead_uses_2015}.
\rev{That said, to protect patient privacy, not all mental health data may be contained within the EHR, and exporting EHR data for VBC may require patient consent \cite{shenoy_safeguarding_2017, leventhal_designing_2015}.}
Clinical data can also be retrieved from \textit{administrative claims databases}, which log diagnostic, treatment, and medication information used to bill healthcare payers \cite{karve_prospective_2009, davis_can_2016}.
Clinics or hospitals may also collect measures of patient satisfaction to understand patients' perceptions of their care \cite{carr-hill_measurement_1992}.

\textit{Active data} require active patient or clinician engagement to be collected, and can be collected with technologies that support digital surveys (eg, smartphones, iPads, computers, \rev{patient portals}) and pen-and-paper questionnaires.
This data include validated self-reported \textit{measures of mental health symptoms}, which quantify symptom presence and/or severity for specific mental health disorders, such as the PHQ-9 for major depressive disorder \cite{kroenke_phq-9_2001}, or the GAD-7 for generalized anxiety disorder \cite{spitzer_brief_2006}.
Active data can also include clinician-rated scales, collected during clinical interviews \cite{andersen_brief_1986}.
Outside of symptoms, self-reported and clinician-rated measures can also quantify \textit{functioning}, as mental health symptoms can impair functioning including cognition, mobility, self-care, and sociality \cite{ustun_measuring_2010}. 
Self-reported measures can also quantify how well patients and their mental health clinicians collaborate towards shared goals, complete tasks, and bond, called \textit{working alliance} \cite{hatcher_development_2006}.
The discussed scales typically quantify persistent symptoms or functional impairment.
Researchers have used everyday devices, such as smartphones, to collect more in-the-moment symptoms via questionnaires called ecological momentary assessments (EMAs) \cite{wang_crosscheck_2016, hsieh_using_2008}.
EMAs can also collect \textit{engagement data}, measuring, for example, medication adherence, or participation in behavioral interventions, such as mindfulness exercises \cite{militello_digital_2022, klasnja_how_2011}.
Active data can be stored in clinical records, like an EHR, but significant investments have not been made to build structured EHR fields for storing active data \cite{pincus_quality_2016}.

In addition to active data, sensors embedded in devices (eg, smartphones, wearables) and online platforms have created opportunities to collect \textit{passive data} -- data collected with little-to-no effort -- on behavior and physiology \cite{nghiem_understanding_2023}.
Passive data can be used to estimate signals related to functioning, including social behaviors, mobility, and sleep \cite{mohr_personal_2017, saeb_relationship_2016, saeb_scalable_2017}, and more recently, researchers have investigated if passive data can measure engagement in therapeutic exercises \cite{evans_using_2024}.
Prior work has also studied whether passive data can estimate symptom severity \cite{adler_measuring_2024, das_swain_semantic_2022, meyerhoff_evaluation_2021, currey_digital_2022}.
The use of passive data in treatment is limited: \rev{while passive data can be collected within EHRs \cite{apple_healthcare_2024, metrohealth_track_2024, pennic_novant_2015}, established clinical guidelines for passive data use in care do not exist, and use is often limited to patients who are motivated to share passive data with their healthcare provider \cite{nghiem_understanding_2023}}.

It is challenging to identify what mental health data are most relevant to HITs in certain contexts, given their variety.
Li et al. proposed a 5-stage model to work through these challenges, specifically in the context of \textit{personal informatics systems}, where users collect data for self-reflection and gaining self-knowledge.
These five stages are preparation, collection, integration, reflection, and action \cite{li_stage-based_2010}.
In this work, we study how HITs can support mental health outcomes data as a part of value-based mental healthcare, inspired by three out of these five stages, specifically \textit{preparation}, understanding what data to collect; \textit{collection}, gathering data; and \textit{action}, how data is used.
We focus on these three stages because they capture existing challenges to design HITs that support VBC, which we review in Section \ref{sec:rw:vbc}.

\subsection{Value-based Mental Healthcare}
\label{sec:rw:vbc}
The World Economic Forum defines \textit{value-based care} (VBC) as a \textit{``patient-centric way to design and manage health systems''} and \textit{``align industry stakeholders around the shared objective of improving health outcomes delivered to patients at a given cost''} \cite{world_economic_forum_value_2017}.
VBC intends to change how healthcare is paid for, away from \textit{fee-for-service} payment models -- where payers reimburse providers for the number of services they provide -- towards paying for services if they deliver ``value'' to the healthcare system \cite{brown_key_2017}.
In practice, VBC is implemented by paying providers a set rate for managing patients' health, sharing savings if specific cost or utilization targets are met, and/or by offering financial incentives for payers and providers based upon \textit{quality measures}, which quantify the ``value'' of care \cite{world_economic_forum_moment_2023, health_care_payment_learning__action_network_alternative_2017}.
These changes shift some of the financial risk of healthcare from payers to providers.
In fee-for-service models, providers continue to be paid as they provide more services.
In VBC, providers may lose money if services cost more than set rates, specific cost/utilization targets are not met, or if care quality suffers \cite{novikov_historical_2018, health_care_payment_learning__action_network_alternative_2017}.

Standardized quality measures guide payers and providers to deliver services that improve health outcomes and reduce cost.
The Donabedian model categorizes quality measures into three areas: (1) \textit{structure} -- the material, human, and organizational resources used in care (eg, the ratio of patients to providers); (2) \textit{process} -- the services provided in care (eg, the percentage of patients receiving immunizations); and (3) \textit{outcomes} -- measuring the effectiveness of care (eg, surgical mortality rates) \cite{donabedian_quality_1988, endeshaw_healthcare_2020,agency_for_healthcare_research_and_quality_types_2015}.
While structure and process measures are more actionable -- hospital systems can hire more staff, or modify care practices -- their relationship to outcomes can be ambiguous \cite{quentin_measuring_2019}. 
In contrast, outcome measures most clearly represent the goals of care, but can be biased by factors outside of providers' direct control, including co-occurring health conditions that complicate treatment success \cite{lilienfeld_why_2013, quentin_measuring_2019}.
To reduce bias, statisticians apply a \textit{risk-adjustment} to outcome measures, using regression to model expected care outcomes observed in real-world data, based upon variables known to moderate treatment effects \cite{lane-fall_outcomes_2013}.
The quality of provided health services for a specific patient can then be determined based upon whether a patient's health outcomes exceed or underperform expectations.

Mental healthcare has faced specific challenges implementing VBC.
Some of these challenges can be attributed to ambiguity on how to design health information technologies (HITs) that store outcomes data tying provided services to value \cite{world_economic_forum_value_2017}.
\textit{Preparation challenges} revolve around identifying standardized outcome metrics to store in HITs.
Current quality monitoring programs incentivize using symptom scales as standardized care outcomes \cite{morden_health_2022}.
Patients often experience a unique constellation of symptoms that cut across multiple disorders (eg, major depressive disorder and generalized anxiety disorder) \cite{boschloo_network_2015, cramer_comorbidity_2010, barkham_routine_2023}, making it difficult to identify a limited set of symptom scales to track outcomes across patients.
Given these challenges, researchers have proposed using other data types as an alternative to symptom scales within VBC \cite{hobbs_knutson_driving_2021, oslin_provider_2019}. 
For example, scholars and healthcare providers have argued that functional and engagement outcomes may be a promising alternative to symptom scales. 
Engagement is the proximal outcome of many mental health treatments, improved functioning is often more important to patients than symptom reduction, and functional outcomes measure treatment progress across patients living with different mental health symptoms or disorders \cite{stewart_can_2017, tauscher_what_2021, pincus_quality_2016}.

In terms of \textit{data collection}, it is estimated that less than 20\% of mental health clinicians practice measurement-based care (MBC) -- the process of collecting, planning, and adjusting treatment based on outcomes data -- specifically symptom scales \cite{zimmerman_why_2008, fortney_tipping_2017}, despite evidence that MBC improves outcomes \cite{barkham_routine_2023}. 
MBC is usually implemented by having patients routinely self-report symptoms during clinical encounters using validated symptom scales, like the PHQ-9 for depression, or the GAD-7 for anxiety \cite{wray_enhancing_2018}.
Mental health clinicians choose to not practice MBC for many reasons. 
Electronic health records (EHRs) often do not have standardized fields to support symptom data collection, clinicians perceive that symptom scale administration disrupts the therapeutic relationship, and clinicians are often not paid to administer symptom scales \cite{lewis_implementing_2019, desimone_impact_2023, oslin_provider_2019}.
These barriers call for work centering mental health providers in designing HITs that effectively engage providers in outcomes data collection.

\textit{Action} challenges stem from both perceptions of how outcomes data could be used in care, and challenges towards attributing accountability for care.
For example, clinicians are often not trained to use outcomes data in care, and worry that they will be held accountable and penalized if outcomes data reveal that their patients are not improving \cite{lewis_implementing_2019, desimone_impact_2023}.
There are also concerns that outcomes data could be gamed: biased reporting that artificially inflates performance metrics \cite{kilbourne_measuring_2018}.
In addition, it is difficult in mental healthcare to attribute accountability to specific actors (eg, specific providers) in care systems.
Mental healthcare is often ``siloed'' from physical healthcare, though both physical and mental health outcomes are strongly intertwined (eg, individuals living with schizophrenia suffer from chronic physical health conditions) \cite{pincus_quality_2016}.
Thus, existing value-based mental healthcare programs may hold both physical and mental health clinicians \textit{jointly accountable} by sharing cost savings across different types of providers \cite{hobbs_knutson_driving_2021}.

Taken together, this prior work demonstrates challenges designing HITs that support value-based mental healthcare.
Integral to the design of these HITs are mental health clinicians, who are asked to participate in outcomes data collection, which clinicians have found challenging, and will be held financially accountable to the outcomes data HITs store.
Given these challenges, this work centers mental health clinicians' perspectives on how to design HITs that support value-based mental healthcare.
By centering clinicians' perspectives, we looked to gain a deeper understanding of their workflows and incentives to adopt HITs, and integrate this knowledge into the design and development of HITs supporting value-based care. 
The following section details the methodology used in this study.
\section{Methods}
\label{sec:methods}

We conducted interviews with mental health clinicians to explore how they would design health information technologies (HITs) that support value-based mental healthcare.
Methodologically, we were inspired by work in speculative design to imagine futures where VBC is mandated, and then brainstorm with participants how HITs could support VBC outcomes data storage, collection, and use \cite{hockenhull_speculative_2021, wong_speculative_2018}. 
In this section, we detail the study procedures, including participant recruitment (Section \ref{sec:methods:participants}), background information (Section \ref{sec:methods:participants-backgrounds}), how data was collected and analyzed (Section \ref{sec:methods:data}), and our positionality (Section \ref{sec:methods:positionality}). 
All study procedures were approved by the coauthors' institutional review board (IRB). 

\subsection{Participant Recruitment}
\label{sec:methods:participants}
We enrolled as participants mental health clinicians, specifically practicing psychiatrists, clinical psychologists, licensed clinical social workers (LCSWs), and licensed mental health counselors (LMHCs).
We intentionally recruited providers from these different clinical orientations to gather different perspectives on designing HITs \cite{mental_health_america_types_2024}. 
Participants were recruited via a combination of convenience, purposive, and snowball sampling \cite{etikan_comparison_2015, goodman_snowball_1961}.
Specifically, a recruitment email and flier were sent to staff working at academic medical centers across the United States. 
\rev{Recruitment emails were often forwarded to providers who worked in smaller, private practices or community health settings, to help us gain perspectives from mental health clinicians working in diverse settings, treating different types of patients.} 
Within the qualitative tradition \cite{braun_one_2021}, our goal for this work was not to gather perspectives representative of mental health clinicians as a whole, but instead to deep dive with our participants into the complexities of designing HITs that support VBC.

\subsection{Participants' Backgrounds}
\label{sec:methods:participants-backgrounds}

\rev{Table \ref{tab:participants} summarizes background information for the 30 mental health clinicians who participated in the study.
This background information was collected during an intake survey, which was administered after participants provided informed consent for our study.
Apart from data collected within this intake survey, we often asked participants during our study interviews to provide background information regarding their current payment arrangements.
Most of our participants took traditional, fee-for-service payments (public and private), or asked their private practice patients to pay for care out-of-pocket.
A few participants (eg, SW28) worked in health systems transitioning to value-based payments.
Many participants were unfamiliar with VBC.
}

\begin{table*}[t]
\begin{tabular}{ll}
\toprule
Number of participants & 30 mental health clinicians \\ 
\midrule
Clinical training           & 13 Clinical Psychology \rev{(CP)} \\
                            & 6 Psychiatry \rev{(PS)} \\    
                            & 8 Clinical Social Work \rev{(SW)} \\
                            & 2 Mental Health Counseling \rev{(MC)} \\
                            & 1 Family and Marriage Therapist \rev{(FT)} \\
\midrule
Practice setting            & 16 Academic Medical Center \\
                            & 14 Private Practice \\
                            & 5 Community Mental Health Center \\
                            & 2 Employee Assistance Program \\
\midrule
Geographic location (in the USA)    & 26 Northeast \\
                                    & 2 Southeast \\
                                    & 2 West Coast \\
\bottomrule
\end{tabular}
\caption{Background information of the study participants. Participants could list multiple practice settings.
\rev{Clinical training abbreviations (eg, ``CP'') are used within Section \ref{sec:findings}.}
}
\label{tab:participants}
\vspace{-5pt}
\end{table*}

\subsection{Data Collection and Analysis}
\label{sec:methods:data}

All participants were asked to provide informed consent after being provided complete information about the study procedures.
Interviews were held via Zoom over two 1-hour sessions attended by the first three authors, and participants were reimbursed \$30 per hour for their time.
The first session was a semi-structured interview where we asked clinicians about their current care practices, specifically how they used data -- defined broadly, collected with or without technology -- in care.
We specifically asked participants about their perspectives on \textit{measurement-based care} (MBC), the practice of collecting and using data in care that would power HITs supporting VBC \cite{kilbourne_measuring_2018}.
We then asked participants further questions about how they used this data to measure care outcomes, how technology was involved in this process, and whether providers were accountable to achieve certain care outcomes.
Interview questions were broad to allow for on-the-spot adaptation and probing \cite{barriball_collecting_1994}.

In the second session, participants completed two design prompts.
These prompts were motivated by work in speculative design \cite{hockenhull_speculative_2021, wong_influence_2008}, to imagine futures where MBC and VBC were mandated and to understand how clinicians would collect and report outcomes data as a part of these programs.
The first prompt asked participating clinicians to imagine a world where they were mandated to use outcomes data as a part of care, and to brainstorm what data they would prioritize.
The second prompt was motivated by the five-star quality rating system used by the United States Center for Medicare \& Medicaid services (CMS) \cite{center_for_medicare__medicaid_services_five-star_2022}.
Participating clinicians were asked to imagine that as a part of VBC, CMS wanted to design ``mental health quality star ratings'' to measure patient outcomes and care quality across clinics and health systems.
Participants were asked to brainstorm what data should be included in this new star rating program.
After responding to each prompt, we discussed with participants the data they included in their responses, and asked probing questions to further understand how HITs could support data storage, collection, and use.
Full interview guides can be found in Appendix \ref{appendix:guide}.

Interviews were recorded with participants' permission, transcribed by a professional service, and de-identified.
Transcripts were analyzed using a reflexive thematic analysis approach adopted from \cite{braun_using_2006}.
This approach combined both inductive and deductive elements.
Codes and themes arose from the data, but were guided by our research interests and the literature \cite{braun_one_2021}, specifically the stages of preparation, collection, and action from Li et al. \cite{li_stage-based_2010}.
The first author qualitatively coded all transcripts.
Codes were iteratively refined, resulting in a final codebook, and all transcripts were recoded using the final codebook.
Themes were developed from the codes by the first author, with support from the second and third authors who also participated in the interviews and validated that the themes represented participants' views.
The codebook used to generate each theme can be found in Appendix \ref{appendix:codebook}.

\subsection{Positionality}
\label{sec:methods:positionality}

The first, second, and third authors are graduate students in computer and information science. 
These authors recruited participants, collected, and analyzed all of the data. 
One author is a clinical researcher and practicing mental health clinician who worked with the first author on the study protocols, and did not participate in the study. 
Another author is a health policy researcher, who is an expert on both digital mental health and value-based care.
The final author is a researcher in computing and information science. 
All authors were based in the United States, and thus our findings and perspectives are greatly informed, and potentially limited by, our knowledge of the United States healthcare system.
\section{Findings}
\label{sec:findings}

Our findings highlight three themes exploring the design of health information technologies (HITs) that support outcomes data preparation, collection, and use within value-based mental healthcare.
(1) With regards to \textit{preparation}, participants preferred that HITs store functional and engagement outcomes for VBC, as compared to symptom scales or other outcomes data, because they believed functional and engagement outcomes were most directly tied to provided mental health services (Section \ref{sec:findings:preparation}).
(2) Participants also perceived that data \textit{collection} could be improved by investing in HITs that support standardized fields to collect mental health outcomes data, and saw opportunities for devices collecting both active and passive data to improve data collection (Section \ref{sec:findings:collection}).
(3) Finally, participants emphasized that \textit{actions} with outcomes data must hold payers, providers, and social services jointly accountable to care outcomes, and outcomes data need to be risk-adjusted, otherwise providers may prioritize easier to treat patients that inflate outcome metrics (Section \ref{sec:findings:action}).
Throughout our findings, participants are referred to with a unique identifier (eg, CP30) to maintain anonymity.
\rev{These identifiers indicate participants' clinical training (CP = Clinical Psychology, PS = Psychiatry, SW = Social Work, MC = Mental Health Counseling, FT = Family and Marriage Therapist, see Table \ref{tab:participants}).}
Participants referred to individuals receiving mental health services as both ``patients'' and ``clients'', and we use these terms interchangeably in our findings.

\subsection{Preparation: What Outcomes Data Should HITs Store?}
\label{sec:findings:preparation}

A foundation of building HITs for value-based mental healthcare are determining the standardized outcomes data these technologies should store.
Participating clinicians recognized the value of standardized outcomes data. 
As SW28 mentioned, \textit{``I think we have to have some concrete thing that's going to say, `You're getting better. This treatment is working.' ''}
But, participants believed it would be challenging to identify a limited set of outcomes data to use for VBC, even for a single patient or within a single disorder.
Participants mentioned how patients often present in care with multiple symptoms co-occurring across disorders, and collecting data to track all of their symptoms was burdensome:

\begin{quote}
    \textit{``You can't ask questions about absolutely everything. Sometimes you find the patient talks about how they're anxious about their parents, their family and their friends. I give them a longer anxiety scale that hits social anxiety, school anxiety, separation anxiety. 
    But I'm being forced to do all of these assessments and I'm not getting a really good reason why other than because you have to.''} (SW38)
\end{quote}

Given these complexities, we weighed with participants what outcomes data they preferred to use within VBC, and identified two themes.
First, drawing upon their clinical experience, participants believed that symptom scales -- for example, self-reported depression scales -- would be difficult to use. Participants described that symptom scales were difficult to interpret across patients, and did not accommodate patients who identify with different cultural backgrounds (Section \ref{sec:findings:preparation:symptoms}).
Instead, participants preferred using a combination of functional and engagement outcomes data that they believed better reflected patients' goals for care, were more closely connected to treatment, and were relevant across patients presenting with different disorders or symptoms (Section \ref{sec:findings:preparation:func-engagement}).
\rev{
By functional data, participants referred to data that quantified a patient's ability to participate in day-to-day life, including their cognition, mobility, ability to work, and maintain healthy relationships.
By engagement, participants referred to patients' engagement in treatment, including their ability to practice skills or behavior change exercises learned in care, take prescribed medication, or make safety plans for harmful (eg, suicidal) behaviors.
Participants mainly imagined forms of data captured within clinical encounters. 
We discuss in Section \ref{sec:findings:collection} participants' perspectives on using data captured both within and outside of clinical encounters for VBC.
}

\subsubsection{Challenges Using Symptom Scales as Outcomes Data in VBC}
\label{sec:findings:preparation:symptoms}

We began our interviews asking participants about using standardized symptom scales as outcomes data, as existing quality metrics and measurement-based care programs advocate for collecting standardized symptom scales \cite{morden_health_2022, jacobs_aligning_2023}.
These scales quantify symptom severity for specific mental health disorders, and include self-reported symptom scales such as the PHQ-9 for major depressive disorder, or GAD-7 for generalized anxiety disorder.
Symptom scores are added together to provide an overall measure of treatment progress, and could be shared with regulators as outcomes data in VBC.
Participants believed symptom scales were useful for communicating patients' diagnoses for \textit{``insurance repayment''} (SW55) because they give \textit{``a common language''} (CP51).
CP51 also believed symptom scales were useful for understanding if \textit{``there are specific clusters of symptom coming together to understand 
if I have an intervention that targets those symptoms''} (CP51).
But, our participants were uncomfortable using symptom scores as outcomes data within VBC, because patients have challenges interpreting and reporting symptoms.
SW58 explained:

\begin{quote}
    \textit{
    ``A score on the PHQ-9 can get worse because of external factors. 
    Somebody loses a job, gets a divorce, their child is sick, these things can happen that make stress or depression feel much harder to deal with.
    But it doesn't mean that the client is getting worse.''} (SW58)
\end{quote}

We further probed participants about factors that distort symptom scores.
For self-reported scales, some participants described internal and environmental factors that affect self-reporting.
SW37 mentioned how patients may have \textit{``a literacy issue and do not fully gather the meaning of all the questions''} or \textit{``do not feel comfortable fully disclosing their answers. We will see a discrepancy sometimes between how they fill out the form with their doctor and how they fill it out as a mental health professional.''}
CP30, a child and adolescent psychologist, stated that some children \textit{``just tend to kind of rate symptoms on the higher end.''}
One participant, a psychiatrist, described their own reporting behaviors to explain why symptom scores are difficult to interpret at face-value:

\begin{quote}
    \textit{``If I were to take a PHQ I would probably score highly on it, not because I'm depressed, but because when the questions say `You spend a lot of days not wanting to get out of bed,' or `You overeat,' and I'm like, `Yeah, I do, but I'm not doing it because I'm depressed, I'm doing it because I am lazy.' 
    The context is important.
    The hard numbers taken out of context aren't fully accurate.''} (PS25)
\end{quote}

Aside from self-reports, our participants also mentioned how it was difficult to interpret clinician-rated symptom scales.
SW38 would give patients \textit{``baseline assessments and my colleagues would be like, `Oh my god, the patient is so depressed. We have to give them this really extreme, very intense treatment.' ''}, but the participant challenged their colleagues to see symptom scores as \textit{``a piece of a whole picture''}.
Participants described how they would cross-reference symptom scales with other providers to improve their understanding of patients.
One participant, who treated patients with emergent psychotic symptoms, stated that they spend \textit{``30 minutes dissecting what patients have said in different providers' offices, trying to figure out if they've crossed the threshold to a first psychotic episode.''} (CP35).
Another participant mentioned that providers would report, for the same patient, different levels of symptoms quite frequently:

\begin{quote}
    \textit{ ``There's been multiple times where I am rating somebody at lower risk and another clinician rates them at higher risk and it's the same day and program. What do you do about that? How do you work? How do you provide the right treatment?''} (SW28)
 \end{quote}

Participants also believed that symptom scales did not accommodate patients from different cultures.
One participant mentioned how \textit{``there is stigma around sharing one's mental health and so in the hospital system where I work, there are people from so many different cultural backgrounds''}, and that symptom scales would be \textit{``aligned and more accurate for people who are open and coming from a cultural background where there's open discussion of mental health and symptomology''} (SW37).
Another participant, described that symptom scales were not developed inclusively, and the \textit{``evidence-based is pretty self-selecting''} (CP34).
We probed this participant further to understand how this might impact using symptom scores as outcomes data:

\begin{quote}
    \textit{``I feel mixed about this because the way we have developed these measures and the people on whom they have been developed for. They're just not always accurate, inclusive, or culturally appropriate. 
    I don't really see a world where they fully capture the clinical picture for somebody.''} (CP34)
\end{quote}

\subsubsection{Participants Preferred using Functional and Engagement Outcomes Data}
\label{sec:findings:preparation:func-engagement}

The prior section describes various challenges participating clinicians saw using symptom scales as outcomes data supporting VBC.
Given this, we asked our participants for their perceptions regarding alternative types of outcomes data that could be used.
Some participants mentioned using measures of patient satisfaction, but perceived that satisfaction could be biased by aspects of care unrelated to health outcomes, such as \textit{``if the patient likes the hospital's food''} (PS23).
We also asked our participants if care utilization data extracted from EHRs -- such as psychiatric hospitalizations -- were useful outcomes data, but participants were wary to create a culture that discourages utilization: \textit{``If I'm being measured on how many of my clients go to the emergency room, I don't care. Not that I don't care, but, if going to the emergency room was the best decision for that client, what am I going to do?''} (SW50).
Other participants brought up working alliance scales -- that measure the patient-clinician relationship -- but described alliance not as an outcome of care, but an important aspect \textit{``at the beginning of care because you're trying to get that buy-in for treatment''} (CP51).

Instead, participants advocated for using a combination of functional and engagement data as outcome measures, and saw these data types as more aligned with patients' care goals \rev{(examples in Table \ref{tab:findings:preparation:func-engagement-data})}. 
CP30 mentioned how impaired functioning was \textit{``why a lot of people seek treatment. They feel like something is messing up their life in some way. Their goal is to be able to go to school, hang out with friends, spend time with family, whatever it is.''}
Another participant, who treated individuals living with obsessive-compulsive disorder (OCD), found that \textit{``symptom relief itself is not terribly motivating for most people. If you are hamstrung by fears of household chemicals, nobody wakes up in the morning and says, `Oh, boy, I can't wait to get used to these household chemicals.' ''} but instead they \textit{``really focus more on functional gains''} and \textit{``my goal is to get somebody out of the house and interacting with friends''} (FT60).

\begin{table*}[t]
\begin{tabular}{l|l}
\toprule
\textbf{Functional data} quantifying a patient's ability & \textbf{Engagement data} showing participation in \\
to participate in everyday life & skills, exercises, or behaviors relevant for care \\
\midrule
\tabitem Attending school & \tabitem Leaving the house, if fearful of doing so \\
\tabitem Maintaining healthy physiological stress & \tabitem Medication adherence  \\
\tabitem Spending time with family and friends & \tabitem Reducing obsessive handwashing  \\
\tabitem Attending work & \tabitem Communicating suicidal ideation safety plans \\
\tabitem Staying physically active & \tabitem Going to the gym \\
\tabitem Consistent eating behavior & \tabitem Eating a meal \\
\bottomrule
\end{tabular}
\caption{\rev{Examples of functional and engagement data from our findings. 
Engagement data (eg, eating a meal) was often described as a proximal outcome of care and functioning a distal outcome (eg, consistent eating behavior).}
}
\label{tab:findings:preparation:func-engagement-data}
\end{table*}

We asked participants to explain why functional improvements were not captured by symptom scales.
In other words, if functioning improves, why should we not expect symptoms to decrease?
Participants explained that symptom reduction was not the singular outcome of treatment, but treatment intends to improve functioning even if symptoms persist. 
For example, PS24, a psychiatrist treating patients living with schizophrenia, mentioned how they \textit{``tend to have very chronic patients where the goal isn't to get rid of symptoms, but the goal might be to make symptoms interfere with their life less''} and their patients may \textit{``have ongoing voices and paranoia, but they've gotten to the point where they're able to ignore the voices and attend work''}
SW58 agreed, stating that \textit{``when I think about somebody that experiences psychosis or bipolar disorder or depression, you may have this for your whole life. If the goal is to have fewer symptoms, am I setting you up to fail from the start?''} and they work with patients to understand \textit{``given what your life is, how do you want to live? Maybe there's specific things, maybe you want to go back to school.''}.

Furthermore, participants believed that functional outcomes were likely to improve if their patients engaged in care, and saw engagement as the most proximal outcome of care.
For example, many of our participants were psychotherapists who asked patients to practice specific skills or change behavior as a part of treatment.
FT60 mentioned how they \textit{``had somebody who was washing their hands a hundred times a day and driving his family nuts with accommodations''} and they had their patient \textit{``use judicial safety behaviors to play with his daughter who's crawling around on the floor and then take a shower afterwards.''}
A few months into treatment the \textit{``patient is still washing his hands a hundred times a day, but he and his family are tons happier than they were. They're raving about how well they're functioning and working together now}'' (FT60).
CP45 mentioned how for patients with \textit{``panic disorder, I'd want to have some behavioral data on what they are avoiding or how frequently they are getting out of the house, depending on the specifics of that person.''}
Another participant mentioned how, by tracking engagement, they might feel more confident that \textit{``someone having passive suicidal thoughts would have no intent to act on them''} because they \textit{``have a supportive family that they communicate to and a safety plan in place''} (CP46).
CP43 saw treatment as successful if patients consistently engaged in care, even if symptoms were not fully reduced:

\begin{quote}
   \textit{``Their [symptom] scores do cut in half, but don't move much beyond that and stay relatively stable. 
   If they practice their skills and those are well-developed, they got what I am aiming to provide for them.
   Sure their scores aren't zero, but that might just be because of their personality, environment, social context.''} (CP43)
\end{quote}

Unlike symptom scales, participants saw functioning and engagement as \textit{``trans-diagnostic''} (CP42), measuring care outcomes across patients experiencing different symptoms or disorders.
FT60 qualified that measuring symptoms were not irrelevant, but called for \textit{``a shift from symptom-focused metrics to patient-focused metrics, which can include the symptoms.''}
CP33 wanted to prioritize engagement outcomes for complex cases, giving an example of \textit{``a patient who had comorbid substance use disorder, PTSD [post-traumatic stress disorder], borderline personality disorder, there's a lot of suicidality, a lot of very, very intense mood, depression and anxiety''} that \textit{``those intense things, really intense urges, really intense depression, that didn't go away''} but the patient \textit{``developed trust and she kept coming to therapy. She missed, maybe, four sessions all year. Those are therapeutic gains. She internalized some hope that progress is possible.''}
CP30 further explained the importance of engagement and functional outcomes across conditions:

\begin{quote}

\textit{``If someone has one depressive episode, they will likely have another episode.
Someone who has generalized anxiety disorder may always be a more anxious person. 
Someone with obsessive-compulsive disorder may always be vulnerable to intrusive thoughts. 
It doesn't mean they've failed treatment if they can tolerate the anxiety or cope with the depression, go to work, get out of bed, shower, do the things you have to do, using the skills you learned in therapy.''} (CP30)
\end{quote}

\subsection{Collection: How Can HITs Support Outcomes Data Collection?}
\label{sec:findings:collection}

Our first set of findings describe that participants preferred HITs prioritize storing functional and engagement data as outcomes in VBC.
We then explored with participating clinicians how this data should be collected.
Our related work suggests that existing mental health data are not collected by clinicians because they perceive scale administration as burdensome, administration takes time away from treatment, and \rev{mental health clinicians are not always trained to use outcomes data in care}.
Participants affirmed that data was not collected, stating that \textit{``it's hard to get the buy-in from clinicians who don't have that initial training if they have no reason to do it''} and \textit{``if you are a clinician who does not care and doesn't have buy-in, it's really easy to let it slide off''} (CP35).
PS25 stated that \textit{`it feels like it's hard to seamlessly integrate scales into a session.''}
Another participant stated that data collection is \textit{``extra work and we're not getting paid for it''} (CP48).

Given these complexities, we asked participants about the barriers they saw towards engaging clinicians in data collection, and how HITs could improve outcomes data collection.
First, our participants described existing challenges using HITs to collect and manage outcomes data (Section \ref{sec:findings:collection:challenges}).
Specifically, they described that current clinical data infrastructure, namely electronic health records (EHRs), were not designed to collect mental health data, and it was difficult to acquire funding to improve data infrastructure.
Participants also believed that to engage clinicians in VBC data collection, any mandated data collection should be client-specific, easy to administer, and relevant to decision-making.
In addition, participants saw opportunities (Section \ref{sec:findings:collection:opportunities}) for active and passive data, collected via devices (eg, smartphones, tablets, wearables), to improve engagement in data collection.
But, participants believed that for passive data to be used as VBC outcomes, there would need to be evidence demonstrating that passive data accurately measures the outcomes of care.
Participants also raised practical challenges towards active and passive data use.
Participants wanted to understand who would pay for data collection devices and clinicians' time spent interpreting data, how this data would be integrated into clinical records, and were concerned that prioritizing data collected via devices could increase inequities in care. 

\subsubsection{Existing Challenges Collecting Outcomes Data}
\label{sec:findings:collection:challenges}

We probed participants to understand current challenges towards engaging clinicians in outcomes data collection.
Participants described challenges with existing HITs, specifically the electronic health records (EHRs) used in clinical settings to store and share patient health information, including outcomes data in VBC.
Many participants explained how EHRs were not built for mental health data collection.
For example, CP35 mentioned that \textit{``in the medical center, there's just so much red tape. I know they're trying to integrate outcome measures into our EHR system, but it's so challenging.''}
Participants often operated outside of the EHRs used by other clinicians in their health system.
Therefore, they would \textit{``transcribe scales in the EHR ourselves''}, so that other care providers could access patients' mental health data, but this was \textit{``so open to human error of typing in the numbers''} and thus \textit{``while I want scales to be written in the discharge summary paperwork other providers are getting, that doesn't always happen''} (CP46).
SW49 had personal experience advocating for investments in mental health data collection infrastructure.
Before becoming a clinician, they had worked as the director of data analytics and research at a health system serving patients living with eating disorders. 
They described: 

\begin{quote}
    \textit{``We had built data infrastructure, we were getting data on a weekly basis for all of our own patients across the system, and we were just starting to really incorporate that data into treatment planning and reporting data at the end of care to various stakeholders.
    I was there for three and a half years, and then my position was eliminated in a merger.
    This is an unfortunate aspect of mental health in particular, and especially in the eating disorder space. 
    The margins are really thin and nobody wants to invest in data.''} (SW49)
\end{quote}

Given these challenges, many of our participants chose to not use EHRs for data collection.
For example, SW49 described how their clinic would use \textit{``analog means (eg, recording and sharing measures on paper), which were terrible, but the analog means were more successful.''}
Some participants, specifically those in private practice, could not afford EHRs, and used other software for collecting and managing data.
CP34 described that they \textit{``have an Excel sheet for every client} and \textit{I just plug scale scores in, and then I graph them over time''}.
Another participant, SW17, practiced a form of psychotherapy, called dialectical behavioral therapy (DBT), that requires patients to fill out detailed ``diary cards'' before each clinical encounter. 
They described how \textit{``for every patient we do an EHR note''} but \textit{``DBT ends up being very detailed. I only put general data in the body of my EHR note. But for the actual diary cards, I give patients a paper binder and I keep the binders in my office in a locked cabinet''} (SW17).
Since clinicians do not enter detailed data into EHRs, the data entered into the EHR were often incomplete, missing important information from clinical encounters.
One participant mentioned how missing data could be harmful:

\begin{quote}
    \textit{``A lot of people get lazy about using the EHR and might just type in their note something general like, `the score was this' but not actually record all the individual answers.
    It's important to know where people scored on specific symptoms. 
    For example, on a depression scale, you want to know, are people scoring really highly on suicidality?''} (PS25)
\end{quote}

The prior paragraphs detail challenges but also the necessity to invest in and design HITs for collecting and managing mental health data, supporting data sharing between clinicians and other stakeholders as a part of VBC.
Aside from improving these HITs, participants also described that they would be more likely to engage in outcomes data collection if data were more effectively tied to care.
For example, SW38 described how mandated data were often not relevant for their patients.
They stated that \textit{``some of the assessments I have to do because we're a community-based clinic, but I don't really want to ask a 15-year-old about their heroin use habits if that's not something that is relevant, but I have to.''}
Another participant worried about the burden of outcomes data collection, stating that \textit{``if CMS were to require this data, it's important to do an audit of clinicians' other paperwork requirements when they consider the cost of adding these measures''} (CP42).
Therefore, to encourage data collection, participants wanted scales to be client-specific, easy to administer, and relevant for decision-making.
CP34 stated that the scales they choose to use are \textit{``a jumping off point for interventions. They take very little time, they probably take 30 seconds at the beginning of every session. And they're just really integrated into each session as a way to check in and give the person ownership over what they feel like is bothering them the most.''}

\subsubsection{Opportunities for Technology to Improve Outcomes Data Collection}
\label{sec:findings:collection:opportunities}

Given these challenges, we brainstormed with participants how functional and engagement outcomes could be patient-specific, easy to administer, and integrated into care.
Participants raised how they used technology to collect patient-specific active data relevant for care.
CP44 stated that they \textit{``had clients FaceTime loved ones for meals''} to demonstrate engagement in care.
Another participant mentioned how \textit{``if a patient's goal was, `I need to exercise more'} their patient \textit{``could take a photo at the gym''} (SW37).
Other participants thought active data collection could be difficult to enforce.
CP34 stated that \textit{``in most of my training it's been hard to even get people into treatment. I did some EMA data collection in grad school with people who were using substances, and it was just really, really challenging.''}

Some participants identified opportunities for passive data to reduce the burden of outcomes data collection.
For example, PS23 described how, for their patients, \textit{``panic is probably the one thing that you can see a lot of for PTSD, where you end up having physiological stress from your illness. A wearable will show an increase in heart rate, an increase in blood pressure, perhaps an increase in sweating, breathing, and respiration rate.''}
Another participant wrote that activity data could be useful \textit{``because whether we're talking about somebody who's depressed or we're talking about somebody where there's some health adherence problems, let's say it's following a cardiac healthy lifestyle or even anxiety, physical activity may be relevant''} (CP45).
PS53 was interested in collecting language data because \textit{``language is what we use to treat and diagnose.''} 

Despite participants' interest in passive data, they did not believe that passive data, at face-value, could be an outcome measure in VBC.
Even though passive data could make data collection easier, participants saw that interpreting passive data could be challenging.
CP42 stated that \textit{``I have no training in interpreting sleep data to know what's normal versus not''}.
Another participant mentioned how \textit{``on the physiological data, I thought about going back to discrepancies between what patients say or how they're behaving with me. I would need to reflect on, either with them or by myself with my supervisors, and say, `What does this data mean?'} (PS53).
Participants also raised privacy concerns with passive data collection, stating \textit{``patients have to be down with the device collecting the data and a clinician seeing all of that data just from a privacy perspective''} (CP42) and for language data that there would be \textit{``some resistance, from clinicians actually more than patients being recorded and things like that. It can be high-liability information''} (PS53).

Another consideration regarding the use of passive data in VBC was validation: that passive measurement tools are able to accurately measure care outcomes.
For example, if there were a VBC functional outcome focused on quantifying sleep improvements, participants wanted assurances that devices could accurately measure sleep.
CP35 mentioned that their clinic chose to use more expensive research-grade actigraphs, versus consumer activity monitors, because they believed that consumer devices were not as well validated.
Specifically, they said that \textit{``we used actigraphs. And not just like your Apple Watch, but well-validated actigraphs, because I learned your smartwatches are not well validated for telling you if you're sleeping when you're supposed to be sleeping''} though \textit{``it would be a lot easier if you could use what somebody is already wearing''} (CP35).
Another participant was unwilling to use passive data, and would prefer using self-reported scales in the absence of rigorous validation:

\begin{quote}
     \textit{``I've been intrigued by the promise and disappointed by the execution of devices. I hear from sleep researchers that unless you're getting a very expensive device that's really closely tracking you, your Apple Watch is not doing a great job estimating how deep is your actual sleep. 
     It's probably capturing general trends. 
     Unless the technology improves, I'd be really okay with just having a self-report.''} (CP43)
\end{quote}

Participants raised other practical challenges towards integrating both active and passive data into VBC.
CP35 raised that the need for payment mechanisms to reimburse for devices and interpreting data, stating that currently \textit{``we didn't bill separately for the watch. The clinic operated at a loss if the patient didn't bring the watch back.''}
Another participant wanted sleep data to be integrated into existing health records, stating \textit{``I would love if a portal integrated sleep data''} (CP42).
Other participants worried that an over-reliance on devices for active and passive data collection would cater to higher resourced individuals who could afford devices and share data.
CP34 stated that \textit{``the quality of the data really does, in my opinion, skew sometimes towards the higher resourced individuals''} and SW28 mentioned that \textit{``there's value in behavioral markers. But, in our setting, which is a hospital, a lot of patients are not going to have an Apple Watch or a smartphone. Patients don't even have WiFi.''}

\subsection{Action: How Should Outcomes Data be Used in VBC?}
\label{sec:findings:action}

The prior section suggests opportunities to invest in HITs, and use active/passive data to improve clinicians' engagement in VBC outcomes data collection.
Once outcomes data has been collected, VBC programs use this data to create financial incentives that hold providers accountable to achieve specific care outcomes.
Participants, generally, recognized the need for more accountability.
One participant explained that \textit{``there are incentives to keep your patient caseload the same when you're in private practice, because it's a lot of work to do intakes, and you get comfortable with the people you see. 
And so, if there's a piece of your reimbursement that's tied to meeting an outcome and then discharging and starting anew, it also holds you more accountable''} (CP35).
Participants also raised that VBC could give patients more control over care decisions.
FT60 stated that \textit{``many providers convince patients that they're failing treatment''} and CP35 continued: \textit{``it's really hard to know, as a consumer, whether or not you're seeing somebody whose skills actually back up what they say. 
Value-based care could help you steer whether or not you go to somebody.''}

In this section, we describe both challenges and opportunities participants' perceived towards using outcomes data in VBC.
First, participants stressed that outcomes data would need to hold providers, healthcare payers, and social service organizations jointly accountable if VBC were to fulfill its promise of improving mental health outcomes (Section \ref{sec:findings:action:accountability}).
Second, participants voiced the need for HITs to implement risk-adjustments to outcomes data, otherwise clinicians may prioritize treating simpler patient cases that inflate care outcomes (Section \ref{sec:findings:action:risk-adjustment}).

\subsubsection{Using Outcomes Data for Joint Accountability in VBC}
\label{sec:findings:action:accountability}

After participants brainstormed what outcomes data HITs should store, we asked them how this data should be used in VBC programs to improve care.
Specifically, we were interested in who should be held accountable: payers, providers, or other entities?
Participating clinicians quickly pointed out that it would be very difficult to attribute the outcomes of care to any one specific entity.
Though providers would love to take credit if outcomes did improve, that was not always possible:

\begin{quote}
    \textit{``I don't really care what's causing the improvement. 
    If that's due to my intervention, great. 
    And if not, still great because they're feeling better. 
    But I think that tying outcomes to something very specific is too complex. 
    There's too many extraneous and confusing variables to ever do that.''} (CP51)
\end{quote}

Participants gave many examples highlighting the need for \textit{joint accountability}, where responsibility for care outcomes is shared across different providers, or external entities that influence mental health.
For example, CP46 worked in an adolescent inpatient psychiatry unit treating patients in crisis. 
In their view, crisis care may not translate into long-term outcomes, explaining that \textit{``patients may feel totally better because they're in the hospital removed from all the stress and problems of life. Once they leave, I suspect their symptoms would increase. One week or month here doesn't solve the patient's way of approaching life''} (CP46).
To solve this challenge, PS23 believed that clinicians providing inpatient and outpatient services -- the long-term care patients receive after discharge from inpatient -- should be held jointly accountable: \textit{``you could look at across the system, but may not be able to look at for the individual provider. From the patients with depression that we treat as inpatients who then go to our outpatient setting, 85\% still meet criteria for remission after one month. That tells us, okay, we as a health system are doing something right.''}
PS53 believed that physical health providers should be accountable for mental health outcomes.
They stated that \textit{``as I'm doing community psych, I'm learning more that outcomes involve physical care, especially if people can't move around, so we need integration''} (PS53).

Outside of holding providers jointly accountable, participants also described how external entities greatly influenced care outcomes.
One participant raised how health insurers should be held accountable, because \textit{``insurers reimburse clinicians so poorly, so the care is not going to be high quality a lot of the time. It's almost like this circular reasoning issue''} (CP34).
Another participant thought that social services, for example housing or education authorities, should also be held accountable for poor mental health outcomes.
They stated that \textit{``in settings like city hospitals, I wish there was more measurement around what interventions have been effective in reducing stressors related to housing, food, and educational access''} (CP34).
Because social factors, like housing, could influence care, some participants described taking a more active role in linking patients to social services.
SW57 mentioned how \textit{``some patients might be looking for mental health housing, so I'll say, `listen, if you want to bring the paperwork into your next session and spend your session with that, we can absolutely do that.' ''}
CP46 and PS23, though, both expressed frustration that social factors could lead to poor care outcomes, and providers, not social services, may be held accountable.
As they expressed:

\begin{quote}
    \textit{``A patient could be in foster care, lose their foster home and become stuck in inpatient care for two months. That outcome, the length of stay, has nothing to do with how much better they were and everything about the systems serving them.}'' (CP46)
\end{quote}

\begin{quote}
    \textit{``We get frustrated when we see a 30-day readmit but then we understand the patient is homeless and it's 30 degrees outside and someone stole their medication.''} (PS23)
\end{quote}

\subsubsection{Risk-Adjusting Outcomes Data to Encourage Fair Treatment}
\label{sec:findings:action:risk-adjustment}

To penalize and reward stakeholders within VBC, participants voiced that the outcomes data HITs share with payers or quality monitoring organizations need to be \textit{risk-adjusted}: adjusting expected care outcomes based upon the difficulty of the patient case.
Otherwise, participants believed that VBC could dis-incentivize clinicians from treating tougher cases to inflate outcome metrics.
CP42 stated \textit{``how long one person needs treatment differs from how long another person needs treatment. And having a strict outcome can mean that people aren't getting enough treatment.''}
Another participant echoed this concern, saying \textit{``it's kind of unfair if you've got someone treating more severe people to be like, `Oh, you suck at your job because you couldn't get your people down to that level' So the challenge is, what do you want your outcome to be?''} (CP43).
FT60 put these terms more starkly:

\begin{quote}
    \textit{``I'm extremely nervous about the impact on care because it's going to turn the clinician against the patient in favor of boosting their scores. 
    I'm fine with outcome measures being exposed to the consumer so that they can make an intelligent decision as to where they want seek care. 
    But I have real concerns about using them for reimbursement criteria or access to care as a consequence.''} (FT60)
\end{quote}

We approached our participants to understand how HITs should perform risk-adjustments to reduce these potential harms.
If outcome metrics were symptom or functional scores, one participant thought about using \textit{``individual changes or change scores rather than absolute zeros. A person who comes in really significantly depressed who moves mild to moderately depressed [on a symptom scale] is a big change. But if you just use absolute scores, that might not reflect that treatment works''} (CP35).
CP42 suggested that change scores should be relative to patients' baselines, stating that \textit{``if one patient's symptom severity were a 10, but they came into me starting at a 24 that would show amazing improvement. Whereas if someone's usually the happiest person in the world and they're now feeling a little depressed, that might be a notable change''} (CP42).

We also asked participants about specific factors risk-adjustment models should account for to estimate expected care outcomes.
PS32 thought models should account for other conditions patients are living with, stating that they can affect mental health outcomes data: \textit{``I have a patient who needed to have bariatric surgery because it's hard to manage their appetite. They have sleep and energy problems that are related to the chronic pain and fibromyalgia. All those other conditions besides depression already bring the PHQ pretty close to a 10, if not higher than a 10}'' (PS32).
Another participant mentioned that models should account for patients' history of mental illness.
Specifically, that they \textit{``tell people starting on a medication that if this is the first time they've been treated for depression or anxiety, I recommend once your symptoms have been alleviated that you stay on the medication for about six months. If they come off the medication at that time there's a 30\% risk of relapse. If this is your second time with an episode of depression or anxiety, the chances of a third relapse are 70\%. The goalposts get moved a little bit''} (PS23).
In addition, PS23 also raised how social factors influence care outcomes, because \textit{``in the last place I worked, 70\% of our patients were homeless. These are people with a high level of needs, a high level of trauma and stress.''}
CP33 agreed, saying that \textit{``I wouldn't expect someone who's coming in with a really severe depression, who has multiple stressors, and maybe less resilience factors, fewer support system, all kinds of things, to come out of that at the same place as someone who has had a relatively supportive and stable household.''}

In addition to using these factors to moderate expectations, participants also believed these factors should moderate the expected length of treatment, because \textit{``I wouldn't necessarily expect progress to happen in the same way or in the same timeframe''} (CP33).
One participant stated that \textit{``obviously we hope that all of our clients improve. Maybe, over a longer period of time we start to see improvement because you had a long period of therapy. But I don't know that timeframe''} (CP44).
Yet, not all participants were convinced that patients need to be in care for a long period of time to see improvement.
As one participant stated:

\begin{quote}
    \textit{``I would expect change. I would 100\% disabuse the notion that you need to be in therapy for years to see progress and instead show that a lot of people can get better from just a few sessions.''} (CP43)
\end{quote}

\section{Discussion}
\label{sec:discussion}
Our findings center mental health clinicians' perspectives on how to develop HITs that support both the goals of value-based care and providers' individual care needs.
In this discussion, we describe the implications of these findings towards future research developing (Section \ref{sec:discussion:tech}) and designing (Section \ref{sec:discussion:design}) HITs supporting value-based care.
\rev{These implications, contextualized with our findings, are summarized in Table \ref{tab:discussion:findings-implications}.}

\begin{table*}[t]
\begin{tabular}{lll}
\toprule
\textbf{Section} & \textbf{Findings} & \textbf{Implications for Future HITs} \\
\midrule
\textbf{Preparation}: what & Symptom scales posed interpretability & Invest in federated HITs that store \\
outcomes data should & and accessibility challenges. & a bundle of symptom, functional, \\
HITs store? & (Section \ref{sec:findings:preparation:symptoms}) & 
and engagement data, allowing \\
\cline{2--2}
& Functional and engagement data were & clinicians and their patients \\
& perceived to be most aligned with & flexibility to monitor VBC \\
& participants' goals for care and VBC. & outcomes most meaningful for care. \\
& (Section \ref{sec:findings:preparation:func-engagement}) & (Section \ref{sec:discussion:tech:selection}) \\
\midrule
\textbf{Collection}: how can HITs & Health systems do not invest in & Support HCI research mapping \\
support outcomes data & standardized data collection &stakeholders' incentives to invest\\
collection? & infrastructure for VBC. & in HITs for mental healthcare, \\
& (Section \ref{sec:findings:collection:challenges}) & and validating how passive \\
\cline{2--3}
& Validated, passive and active measures & and active data can measure \\ 
& could support data collection, but & context-specific functional \\
& data interpretation must be reimbursed. & and engagement outcomes. \\
& (Section \ref{sec:findings:collection:opportunities}) & (Sections \ref{sec:discussion:tech:validating} and \ref{sec:discussion:design:design}) \\
\midrule
\textbf{Action}: how should & Outcomes data should hold providers, & Design services that link outcomes \\ 
outcomes data be used? & payers, and social services jointly & data across providers, payers, and \\
& accountable for care outcomes. & social services in joint accountability \\
& (Section \ref{sec:findings:action:accountability}) & programs. Create methods to \\
\cline{2--2}
& Outcomes data need to be & risk-adjust data, and triangulate \\
& risk-adjusted for physical, mental, &  passive and active data to create  \\
& and social factors that affect treatment. & more robust outcome metrics. \\
& (Section \ref{sec:findings:action:risk-adjustment}) & (Sections \ref{sec:discussion:tech:risk-adjustment} and \ref{sec:discussion:design:design}) \\
\bottomrule
\end{tabular}
\caption{\rev{Summary of our findings and their implications for developing future Health Information Technologies (HITs) that support value-based care (VBC).}}
\label{tab:discussion:findings-implications}
\end{table*}

\subsection{Developing HITs that Support Value-based Mental Healthcare}
\label{sec:discussion:tech}

\subsubsection{\rev{Developing Data Infrastructure}}
\label{sec:discussion:tech:selection}
\rev{
Our findings in Section \ref{sec:findings:preparation} advocate for developing HITs for VBC that store a suite of symptom, functional, and engagement data.
Within this suite of outcomes data, how do we distill what data is most useful?
VBC programs could allow for a \textit{bundle of interlinked outcomes data}: a set of data types validated for VBC, of which clinicians and patients can select a subset to monitor based upon individual care needs.
This ``bundle of data'' is motivated by the concept of a \textit{personalized data pipeline} from personal informatics literature, where individuals have control of what health data they collect and monitor, as well  as how this data is analyzed based upon their specific goals \cite{wiese_evolving_2017, costa_figueiredo_self-tracking_2017, kim_dataselfie_2019}.
The data used for VBC could also evolve over time as care needs change \cite{clawson_no_2015, adler_beyond_2024, sefidgar_migrainetracker_2024}.
For example, a standardized symptom scale could be used at the beginning of care for screening, diagnosis, and treatment selection.
Engagement and functional outcomes could then be used to monitor progress in treatment.
To be specific, a clinician may administer a Y-BOCS -- a standardized symptom scale -- to screen a patient with suspected obsessive-compulsive disorder (OCD) \cite{goodman_yale-brown_1989}.
Then, using examples from Table \ref{tab:findings:preparation:func-engagement-data}, a clinician could engage a patient in behavior change exercises to reduce obsessive behaviors, like handwashing.
Exercise engagement could be monitored by having a patient self-report how often they wash their hands, or by using a wearable device that passively senses handwashing.
The objective of reduced handwashing could be to improve a specific functional outcome important to a patient, such as their ability to spend time with family and friends.
This functional outcome could be tracked using selected questions from a functional measure, like the WHODAS \cite{ustun_measuring_2010}, or using passive data on phone use and communication.
This example shows how a bundle of data could give patients and clinicians the flexibility to develop personalized VBC outcome data pipelines that are most meaningful for care.
}

\rev{
Yet, if clinics collect different data, how do we enable a standardized data sharing infrastructure?
Individual clinics may choose to collect certain types of outcomes data for VBC based upon device constraints, scale administration infrastructure, or patients served. 
In addition, patients' data sharing preferences may limit what data can be extracted from EHRs for VBC \cite{shenoy_safeguarding_2017, leventhal_designing_2015}.
From a technology perspective, this calls for building \textit{federated HITs}, where data is securely collected and stored locally at a hospital or clinic, and VBC program administrators (eg, health insurers, the government) only access specific data types through a virtual repository, with patients' permission \cite{lin_developing_2014, bellika_properties_2007}.}

\rev{From a sociotechnical perspective, clinicians and health systems should better engage patients in what data is being shared with program administrators, and the benefits of collecting and sharing data for VBC \cite{kruzan_perceived_2023}.
Processes such as \textit{dynamic consent} could engage patients on what data is being shared and with whom, as collected data types change over the course of care, or health insurance coverage changes impact who data is shared with \cite{nghiem_understanding_2023, kaye_dynamic_2015}.
This calls for continued research on \textit{consentful interfaces} that better elicit patients' preferences on health data sharing and use \cite{tseng_data_2024, murnane_personal_2018, im_yes_2021}.
By negotiating data use, patients may be more comfortable participating in VBC data collection despite the additional data work.
In addition, this data collection could empower patients with data to show their clinicians what care decisions are working or not working.
Further research with patients is needed to explore preferences for data sharing and use as a part of VBC.
}

\subsubsection{\rev{Developing Passive and Active Care Outcomes}}
\label{sec:discussion:tech:validating}
\rev{Our findings in Section \ref{sec:findings:collection:opportunities} suggest collecting symptom, functional, and engagement data both passively and actively.}
A decade of research in human-computer interaction, ubiquitous computing, and digital mental health has studied how a combination of active and passive data can be used to measure behavioral and physiological signals associated with symptoms of mental illness.
This research has focused on conditions including depression \cite{xu_globem_2023, adler_identifying_2021, nepal_moodcapture_2024}, anxiety \cite{das_swain_semantic_2022}, bipolar disorder \cite{frost_supporting_2013}, and schizophrenia \cite{wang_crosscheck_2016, wang_predicting_2017}.
\rev{Our findings advocate for continuing this behavioral tracking work, centering how passive and active data can measure functional and engagement outcomes contextualized to specific interventions in care.}
For example, Evans et al. recently explored how passive data can measure engagement in therapeutic exercises for treating post-traumatic stress disorder \cite{evans_using_2024}.
Many other clinical interventions involve engaging in behaviors that could also be measured with a combination of active and passive data, including medication adherence, reducing avoidance behaviors \cite{jacobson_behavioral_2001, hopko_behavioral_2004}, or regulating sleep and wake cycles \cite{frank_interpersonal_2001}.

One challenge that arose from our findings is that participants preferred using more expensive, research-grade devices to measure behaviors related to functioning (Section \ref{sec:findings:collection:opportunities}), for example, fine-grained sleep.
Participants believed that research-grade devices were more rigorously validated than less-expensive and more ubiquitous consumer devices.
From an equity perspective, absent reimbursement mechanisms that pay for research-grade devices in care, researchers and companies could prioritize publishing and disseminating data validating that lower-cost, consumer devices accurately measure fine-grained behaviors associated with functioning.
\rev{In addition, VBC programs should be careful on mandating device-driven data collection, which may be difficult for specific populations (eg, homeless) that cannot easily access devices for care.}

Furthermore, future research could focus on evaluating how passive and active data change as individuals' engage in care.
HCI and digital mental health researchers studying active and passive mental health measures often focus on non-clinical populations, for example, students \cite{wang_studentlife_2014, nepal_capturing_2024, nepal_moodcapture_2024} or information workers \cite{nepal_workplace_2023, das_swain_sensible_2024}.
In addition, administrative claims, which track prescriptions and treatments billed to payers, could also track engagement in care, though researchers need to validate that billed claims measure actual engagement (eg, a prescription could be billed even though a patient does not take their medication) \cite{leslie_calculating_2008, yan_medication_2018}.
Thus, future work designing HITs supporting VBC could collect a suite of clinical, active, and passive data using EHRs, claims databases, research-grade and consumer devices, and quantify expectations for how this data changes for different types of patients as they engage in specific clinical interventions.

\subsubsection{\rev{Developing Risk-Adjustment Methods}}
\label{sec:discussion:tech:risk-adjustment}
Finally, our findings in Section \ref{sec:findings:action:risk-adjustment} describe that participating clinicians wanted HITs to \textit{risk-adjust} outcomes data used in VBC.
Otherwise, participants were concerned that providers could ``game'' outcomes by prioritizing simpler cases.
These concerns are valid: a working paper from the National Bureau of Economic Research suggests that pay for performance programs encourage providers to not treat high risk dialysis patients \cite{bertuzzi_gaming_2023}.
Though risk-adjustment could reduce gaming, other scholars argue that risk-adjustment could increase inequities in care by normalizing inferior treatment outcomes for more difficult cases \cite{jacobs_cms_2023}.
We see opportunities for HCI research to work closely with patients, providers, and health economists to develop risk-adjustment methods that do not increase inequities.
For example, metrics measuring the fairness of machine learning and AI models, such as equal opportunity, odds, or demographic parity \cite{hardt_equality_2016, kallus_fairness_2019, kleinberg_inherent_2016} could inform risk-adjustment models by quantifying differences in expected treatment outcomes across sensitive groups (eg, race, gender).
\rev{In addition, expected treatment outcomes would need to be developed for different outcome data ``bundles'', resulting in risk-adjustment paradigms for the different passive and active metrics that may be used in care.}

\rev{
Finally, investing in measurement-based care training and ongoing consultation programs for clinicians \cite{marriott_taking_2023}, or developing methods to triangulate rating scales with, for example, passive behavioral data, may result in VBC metrics that are more robust to variable clinician rating practices.
These triangulation methods could help resolve discrepancies between how patients or clinicians rate symptoms in care and observable behavior.
For example, HCI researchers have imagined how passive measures could act as \textit{digital collateral} -- reifying clinician or patient ratings to get a more complete picture of treatment progression \cite{ernala_methodological_2019, fisher_beyond_2017}.
Accumulating, de-identifying, and sharing outcomes data across clinics for research could enable triangulation methods to create more robust quality metrics across data types and rating practices \cite{adler_call_2022}.
}

\subsection{Taking a Stakeholder-Centered Perspective to Design HITs Supporting Health Systems}
\label{sec:discussion:design}

\subsubsection{\rev{Accounting for Health System-Level Design Challenges}}
\label{sec:discussion:design:challenges}
In this work, we were confronted with health system-level design challenges for HITs.
First, our findings in Section \ref{sec:findings:collection:challenges} describe challenges participants encountered funding HITs that support mental health data collection, which could be solved if providers, health system administrators, and payers were incentivized to invest in and create sustainable funding streams for these HITs.
\rev{Mental healthcare in the United States remains underfunded compared to physical healthcare, despite parity laws, and many mental health specialty care providers do not take health insurance \cite{kannarkat_advancing_2024, rafla-yuan_mental_2024,bishop_acceptance_2014}.}
In addition, mental health has lagged behind other specialties in implementing HITs \cite{kilbourne_measuring_2018}.
Technology adoption incentives have often excluded mental health providers.
For example, the HITECH Act in the United States offered financial incentives to providers that implemented EHRs, but excluded nonphysician providers, including the clinical psychologists and social workers that make up a large part of the mental health workforce \cite{maulik_roadmap_2020, ranallo_behavioral_2016, american_psychological_association_hitech_2012, bureau_of_health_workforce_behavioral_2023}.
\rev{In addition, EHR implementations are expensive, inhibiting smaller provider practices from adopting EHRs \cite{rao_electronic_2011}.}
This may partially explain why many participants in our study -- \rev{many of whom were clinical social workers and psychologists working in small practices} -- did not use HITs, or why the HITs they used did not contain standardized fields to store mental health outcomes data.

Second, participants described that VBC programs should enforce joint accountability (Section \ref{sec:findings:action:accountability}), where payers, providers, and social services coordinate care and are all held financially accountable to care outcomes.
Current approaches towards joint accountability are not straightforward.
In the United States, different organizations rate the quality of care for different stakeholders involved in health service delivery.
For example, the NCQA publishes quality data used to accredit specific health insurance plans \cite{ncqa_health_2024}, while CMS publishes quality data on healthcare providers \cite{cms_quality_2024}.
Though health plans and providers are rated independently, their ratings are intertwined: health plan ratings are affected by the outcomes of service providers, and providers' outcomes are limited by services health plans cover.
Government social services may also be impacted by low quality care. 
Poor social services, leading to a lack of, for example, housing and employment opportunities, can worsen health \cite{bambra_tackling_2010}.
This leads to higher acute healthcare utilization, particularly among individuals receiving government health insurance -- which in the U.S. are primarily elderly, lower income, or individuals on disability \cite{keisler-starkey_health_2023} -- increasing public healthcare expenditures.

More direct approaches to implementing joint accountability would reward or penalize all providers, payers, and social services as population health improves or worsens.
This would force these stakeholders to coordinate care to improve health outcomes.
One example of more direct joint accountability was the Hennepin Health Accountable Care Organization (ACO), a health insurance plan that coordinated and shared healthcare cost savings across different organizations providing social and mental health services \cite{vickery_changes_2020}.
\rev{Enrollees in the ACO had more consistent primary and mental healthcare utilization and improved quality of life \cite{vickery_changes_2020}, while also demonstrating some, though non-significant, cost savings \cite{vickery_integrated_2020}.} 
Yet, health economists have warned that forcing health systems to assume responsibility for social services could disproportionately burden low-resourced health systems \cite{glied_health_2023}. 
But, if risk was shared across health systems, higher-resourced health systems would be incentivized to invest in communities served by lower-resourced health systems, alleviating some of this burden.

\subsubsection{\rev{Designing for Health Systems in HCI}}
\label{sec:discussion:design:design}
\rev{We see these challenges as opportunities for HCI research that critically engages with how funding and health system-level data shape the design of HITs.}
HCI scholars have advocated that HCI focus on aspects of financing and stakeholder coordination that influence the effectiveness of HITs \cite{blandford_hci_2019, colusso_translational_2019}. 
\rev{
For example, the challenges we have identified could be framed as a \textit{goal misalignment} challenge \cite{kirchner_they_2021, schroeder_examining_2020}, where the data collection goals for VBC program administrators -- monitoring health system-level outcomes and costs -- are not always aligned with the goals of patients and clinicians, to monitor individual-level progress in care.
Personal informatics researchers have shown that individuals are more likely to engage in data collection when it aligns with their specific goals \cite{epstein_lived_2015}, and designing technologies to account for the diversity of users' goals can improve the data collection experience \cite{sefidgar_migrainetracker_2024}.
If patients and clinicians can personalize or negotiate collected outcomes data with VBC program administrators, we may be able to better align system- and individual-level data collection aims \cite{gulotta_fostering_2016}.
}

To align goals, Forlizzi argues that HCI take a service-oriented approach to design, which they call \textit{stakeholder-centered design}, to account for the needs of stakeholders and their interactions with technologies \cite{forlizzi_moving_2018}. 
In this work, we centered one specific stakeholder: mental health clinicians.
Future work taking a stakeholder-centered design perspective could uncover how interactions between patients, payers, providers, and social service organizations contribute to the financing and effective use of HITs in VBC. 
Methods from \textit{service design} could center each stakeholder's design requirements for HITs, and how HITs can support interactions within joint accountability programs \cite{hwang_societal-scale_2024, forlizzi_promoting_2013}.
For example, \textit{stakeholder mapping} could map the power dynamics of stakeholders involved in financing HITs \cite{newcombe_client_2003}.
\textit{Service blueprinting} could then identify interactions with HITs across stakeholders in joint accountability programs.
Surfacing multi-stakeholder perspectives are essential towards understanding why a specific HIT may or may not be funded or adopted.
Such methods would allow technology designers to infuse stakeholders' perspectives into the initial development of HITs, build technologies that effectively support VBC, and improve care outcomes for \textit{all} patients.

\subsection{Limitations}

These findings reflect our interpretation of the literature joined with the perspectives of 30 mental health clinicians.
They should not be interpreted to reflect mental health clinicians as a whole.
In addition, we only interviewed mental health clinicians as participants.
A broader perspective of value-based mental healthcare would include the views of other stakeholders, including but not limited to patients, health system administrators, payers, and social service administrators.
We plan to include these perspectives in future work.
\rev{Participants and the authors were based in the United States.
Findings and implications are thus biased towards a U.S. perspective.
In addition, the majority of participants worked in academic medical centers, which are not representative medical centers and clinics across the United States \cite{fisher_academic_2019}.
We often asked participants during the interviews about their current payment arrangements (eg, insurance, out-of-pocket pay), but structured data to support our findings were not collected.}

\subsection{Conclusion}

Our findings illuminate opportunities to design health information technologies that support value-based mental healthcare.
\rev{
Specifically, our findings advocate for flexibility in HIT development, allowing healthcare providers and patients choice to collect and store outcomes data for VBC most aligned with their specific care goals.
Simultaneously, future HCI research could take a more health systems-level perspective towards HIT design, by engaging the multiple stakeholders involved in VBC who influence the effectiveness of these technologies.
}
We hope these findings chart a path towards developing technologies that effectively support healthcare payment programs, clinicians' practice, improve health service delivery, and patient outcomes.

\begin{acks}
D.A. is supported by a National Science Foundation Graduate Research Fellowship under Grant No. DGE-2139899, a Digital Life Initiative Doctoral Fellowship, and a Siegel PiTech PhD Impact Fellowship.
Transcription and participant reimbursement costs were supported by a multi-investigator seed grant through the Cornell Academic Integration Program, awarded to T.C.
Publication costs were funded by a National Science Foundation Grant No. 2212351 awarded to T.C.
A.V.M is supported by NIMH Grant No. K23MH120505.
Any opinions, findings, and conclusions or recommendations expressed in this material are those of the authors and do not reflect the views of the funders.
\end{acks}

\bibliographystyle{ACM-Reference-Format}
\bibliography{references}

\appendix
\clearpage
\section{Interview Guide}
\label{appendix:guide}

\subsection{Session 1: Formative Interview}

\begin{enumerate}
    \item Can you describe your practice, including, but not limited to, the types of patients you see, the conditions you treat, and the treatments you offer?
    \item In this study we are interested in learning about how mental health specialty care providers use or do not use measurements during routine patient encounters to monitor patient progress and inform treatment, also known as measurement-based care, or MBC for short. What do you think about MBC? 
    \item Do you use measurements to inform your practice? If so...
    \begin{enumerate}
        \item Why do you practice measurement-based care? Is it mandated?
        \item What measures do you use? For what patients specific patients, or specific treatments?
        \item How do you collect these measures? 
        \begin{enumerate}
            \item Pen and paper? Is technology involved? 
            \item Do you use patient self-report?
            \item What informs these data collection decisions?
        \end{enumerate}
        \item How do these measures inform your practice?
        \begin{enumerate}
            \item Do you review measures with your patients?
            \item Is technology involved?
        \end{enumerate}
        \item Do you find MBC helpful? Not helpful? Why? What barriers do you face towards using MBC to improve care?
    \end{enumerate}
    \item If you do not use measurement-based care, why do you not use MBC?
    \begin{enumerate}
        \item Have you considered using MBC in your practice?
        \item Are there other practices you use that are like MBC?
    \end{enumerate}
    \item How do you think about measuring outcomes as a part of care? 
    \begin{enumerate}
        \item How do you measure that a patient is improving or deteriorating?
        \item How is technology involved/not involved in this process?
        \item Do your patients ``graduate'' or ``complete'' treatment? What data informs treatment completion or graduation?
    \end{enumerate}
    \item Do others (eg, administrators) in your organization review this information. If so, what do they do with this information?
    \begin{enumerate}
        \item Is there any systems-level oversight?
        \item Evaluation of patient progress?
        \item Accountability to achieve certain care outcomes?
        \item Does technology facilitate this process? Is there technology involved in care coordination, or sharing data for oversight? Why or why not?
    \end{enumerate}

\end{enumerate}

\subsection{Session 2: Brainstorming Session}

 In this session, we will have you brainstorm responses to two prompts on an online slide deck. We will complete each prompt one at a time. After you respond to each prompt, we will review your responses and ask you follow-up questions. Do you have any questions before we begin?

\begin{enumerate}
    \item We want you to imagine that your healthcare system has asked you to start monitoring patient outcomes as a part of your treatment, in order to maximize the care you deliver for your patients. 
    Your department chair, or a supervisor, may review these outcome measures with you at routine check-ins. 
    Please brainstorm a list of measures that you would use to monitor and maximize treatment outcomes for the patients you are treating. 
    Think about measures you would collect and monitor across routine patient encounters. 
    These measures can include both measures you currently collect in your practice, or measures you do not collect, but think would be helpful for monitoring and maximizing patient outcomes. 
    Feel free to brainstorm a list of measures, or draw diagrams or visuals, whatever is most helpful to communicate your ideas. 
    Once you are done, we will have a conversation about what you came up with.
    [Provide example measures, if helpful, such as:]
    \begin{enumerate}
        \item Standardized symptom measures (eg, patient reported outcome measures, like the PHQ-9, GAD-7, etc.)
        \item Quality of life measures
        \item Psychosocial measures
        \item Health behaviors
        \item Idiographic measures, focusing on specific patient goals
        \item Patient satisfaction/experience with treatment
        \item Working alliance
        \item Measures of comorbid conditions (connecting physical and mental health)
        \item Psychiatric events (eg, hospitalization)
    \end{enumerate}
    [Can also discuss:]
    \begin{enumerate}
        \item When would these measures be taken? At the beginning, during, or after the visit?
        \item At what frequency should these measures be taken?
        \item How would these measures be recorded? Self-reported by a patient? Rated by a clinician?
        \item Where would these measures be stored? In the EHR? Another system?
    \end{enumerate}

    \item We want you to now imagine that as part of a push for value-based care, the Center for Medicare \& Medicaid Services, or CMS, are creating “mental health quality star ratings” to measure patient outcomes and care quality across clinics and health systems. 
    CMS wants your help, and is seeking feedback from you, actual clinicians treating patients, to understand how these measures can be designed to most accurately reflect patient outcomes. 
    These ratings may be partially based upon measures that you, the clinician, gather during routine clinical encounters with your patient. 
    In this session, you will have an opportunity to shape how these mental health star ratings are created.
    Please brainstorm measures that your clinic, or department, would use to monitor patient outcomes across your entire clinic/department. 
    These measures would be shared with CMS. 
    Feel free to brainstorm a list of measures, or draw diagrams or visuals, whatever is most helpful to communicate your ideas. Once you are done, we will have a conversation about what you came up with.
\end{enumerate}

This concludes the study. Thank you for participating.
\clearpage
\section{Codebook}
\label{appendix:codebook}
\begin{table}[!htbp]
\begin{tabular}{l|l}
\toprule
\textbf{Theme} & \textbf{Codes} \\
\midrule
& Adherence, engagement as outcomes data \\
& Behavior and physiology as outcomes data \\
& Challenges patients face during measure administration \\
& Interpreting outcomes data \\
& Patient-clinician relationship as outcomes data \\
Preparation: what outcomes data should HITs store?  & Patient goals as outcomes data \\
& Patient satisfaction as outcomes data \\
& Quality of life and functioning as outcomes data \\
& Standardizing outcomes data \\
& Symptoms as outcomes data \\
& Types of patients participant treats \\
& Using measures to understand symptoms \\
\midrule
& Challenges clinicians face administering and using measures \\
& Clinician mental health as outcomes data \\
& Collaborating to create outcomes data \\
& Health information technology access \\
& Interfaces clinicians use to collaborate on outcomes data \\
Collection: how can HITs support outcomes data collection? & Interfaces for viewing trends \\
& Interfaces patient uses to record outcomes data \\
& Passive data as outcomes data \\
& Payers and regulatory bodies mandating outcomes data collection \\
& Pen and paper interfaces for recording outcomes \\
& Utility for clinician to collect outcomes data \\
\midrule
& Accountability to outcomes data \\
& Financial incentives for achieving outcomes \\
Action: how should outcomes data be used in VBC? & Graduating from treatment \\
& How are clinicians paid \\
& Risk-adjusting outcomes data \\
& Social services patients' use \\
\bottomrule
\end{tabular}
\caption{Codes used to generate each theme.}
\label{tab:appendix:codes}
\end{table}

\end{document}